\documentclass[a4paper,superscriptaddress,citeautoscript,twocolumn,prl]{revtex4-1}

\usepackage[T1]{fontenc}
\usepackage{times}
\usepackage{amsmath,amssymb}
\usepackage[pdftex]{graphicx,color}
\usepackage{epstopdf}
\usepackage[a4paper,body={18cm,25.5cm},top=2cm]{geometry}

\setcitestyle{super}

\newcommand*{\citen}[1]{%
  \begingroup
    \romannumeral-`\x 
    \setcitestyle{numbers}%
    \cite{#1}%
  \endgroup
}

\makeatletter
\let\bibsection\rtx@bibsection%
\def\fnum@figure{\textbf{Figure~\thefigure} $|$\ \@gobble}
\makeatother

\begin{document}

\title{Engineered discreteness enables observation and control of chimera-like states in a system with local coupling}

\author{Alexander U. Nielsen}
\author{Yiqing Xu}
\affiliation{The Dodd-Walls Centre for Photonic and Quantum Technologies, New Zealand}
\affiliation{Physics Department, The University of Auckland, Private Bag 92019, Auckland 1142, New Zealand}
\author{Michel Ferr\'e}
\author{Marcel G. Clerc}
\affiliation{Departamento de F\'isica and Millenium Institute for Research in Optics, Facultad de Ciencias F\'isicas y
Matem\'aticas, Universidad de Chile, Casilla 487-3, Santiago, Chile}
\author{St\'ephane Coen}
\author{Stuart G. Murdoch}
\author{Miro Erkintalo}
\altaffiliation{corresponding author: m.erkintalo@auckland.ac.nz}
\affiliation{The Dodd-Walls Centre for Photonic and Quantum Technologies, New Zealand}
\affiliation{Physics Department, The University of Auckland, Private Bag 92019, Auckland 1142, New Zealand}



\begin{abstract}
  \noindent Chimera states -- named after the mythical beast with a lion's head, a goat's body, and a dragon's tail -- correspond to spatiotemporal patterns characterised by the coexistence of coherent and incoherent domains in coupled systems~\cite{abrams_chimera_2004, panaggio_chimera_2015}.  They were first identified in 2002 in theoretical studies of spatially extended networks of Stuart-Landau oscillators~\cite{kuramoto_coexistence_2002}, and have been subject to extensive theoretical~\cite{omelchenko_chimera_2008,laing_dynamics_2009, wolfrum_chimera_2011, maistrenko_cascades_2014, laing_chimeras_2015, clerc_chimera-type_2016} and experimental~\cite{hagerstrom_experimental_2012, tinsley_chimera_2012,larger_virtual_2013,  martens_chimera_2013, viktorov_coherence_2014, larger_laser_2015, totz_spiral_2018, uy_optical_2019} research ever since. While initially considered peculiar to networks with weak nonlocal coupling, recent theoretical studies have predicted that chimera-like states can emerge even in systems with purely local coupling~\cite{laing_chimeras_2015, clerc_chimera-type_2016}. Here we report on experimental observations of chimera-like states in a system with local coupling -- a coherently-driven Kerr nonlinear optical resonator~\cite{leo_temporal_2010}. We show that artificially engineered discreteness -- realised by suitably modulating the coherent driving field -- allows for the nonlinear localisation of spatiotemporal complexity, and we demonstrate unprecedented control over the existence, characteristics, and dynamics of the resulting chimera-like states. Moreover, we show evidence that ultrafast time lens imaging~\cite{suret_single-shot_2016, narhi_real-time_2016, li_panoramic-reconstruction_2017, Ryczkowski_real-time_2018, copie_observation_2018} allows for the chimeras' picosecond-scale internal structure to be resolved in real time.
\end{abstract}

\maketitle

\section{Introduction}

Synchronization of coupled oscillators is a fascinating phenomenon that resonates across nature, technology, and society, with famous manifestations ranging from flashing fireflies to the collective behaviour of applauding audiences~\cite{pikovsky_synchronization:_2003}. Equally fascinating is the fact that, under particular conditions, a network of identical oscillators can split into two spatially distinct domains: one where the oscillators exhibit coherent phase \mbox{synchronization} and another where the phases drift incoherently. The resulting hybrid states, first identified in 2002 by Kuramoto and Battogtokh\cite{kuramoto_coexistence_2002}, have come to be known as chimera states~\cite{abrams_chimera_2004} and inspired a burgeoning field of research~\cite{omelchenko_chimera_2008, laing_dynamics_2009, wolfrum_chimera_2011, maistrenko_cascades_2014, laing_chimeras_2015, clerc_chimera-type_2016,hagerstrom_experimental_2012, tinsley_chimera_2012, larger_virtual_2013, martens_chimera_2013, viktorov_coherence_2014, larger_laser_2015, totz_spiral_2018, uy_optical_2019}.

Chimeras were first identified in studies of coupled oscillators, but similar coexistence of coherent and incoherent domains has also been identified for a range of other systems; the term chimera-like state has been coined to generalize the concept to coupled (discrete) or extended (continuous) dynamical systems~\cite{clerc_chimera-type_2016}. Studies have shown that chimeras and chimera-like states are ubiquitous, manifesting themselves in a variety of nonlinear systems, including human brain networks~\cite{bansal_cognitive_2019}. Experimental observations have been reported, e.g., in chemical~\cite{tinsley_chimera_2012, totz_spiral_2018}, optical~\cite{hagerstrom_experimental_2012, viktorov_coherence_2014, larger_laser_2015, uy_optical_2019}, electronic~\cite{larger_virtual_2013}, and mechanical~\cite{martens_chimera_2013} systems.

While it was initially thought that weakly nonlocal coupling is key to chimera existence, it was later theoretically unveiled that they could arise even in networks subject to purely global~\cite{sethia_chimera_2014} or local coupling~\cite{laing_chimeras_2015, clerc_chimera-type_2016}. In the limit of locally coupled networks, the emergence of chimera and chimera-like states was explained as the pinning of fronts between homogeneous (coherent) and spatiotemporally chaotic (incoherent) domains, facilitated by the intrinsic discreteness of the networks under study~\cite{clerc_chimera-type_2016}. As of yet, however, no experimental demonstrations of such dynamics have been reported to the best of our knowledge: all experimental studies of chimeras have been realised in systems with nonlocal coupling. Moreover, robust protocols of individual chimera creation, annihilation, and control have remained elusive. Here we report on experimental observations of chimera-like states in a continuous, locally coupled system, where front pinning is enabled through artificially engineered discreteness. In addition, we demonstrate unprecedented control over the chimeras creation and annihilation, showing that they can be individually addressed, i.e., turned on and off at will. Finally, we show evidence that ultrafast time lens imaging~\cite{suret_single-shot_2016, narhi_real-time_2016, li_panoramic-reconstruction_2017, Ryczkowski_real-time_2018, copie_observation_2018} allows for the direct and real-time examination of the chimeras' picosecond-scale, fluctuating internal structure.

\begin{figure*}[!t]
\centering
  \includegraphics[width = \textwidth, clip = true]{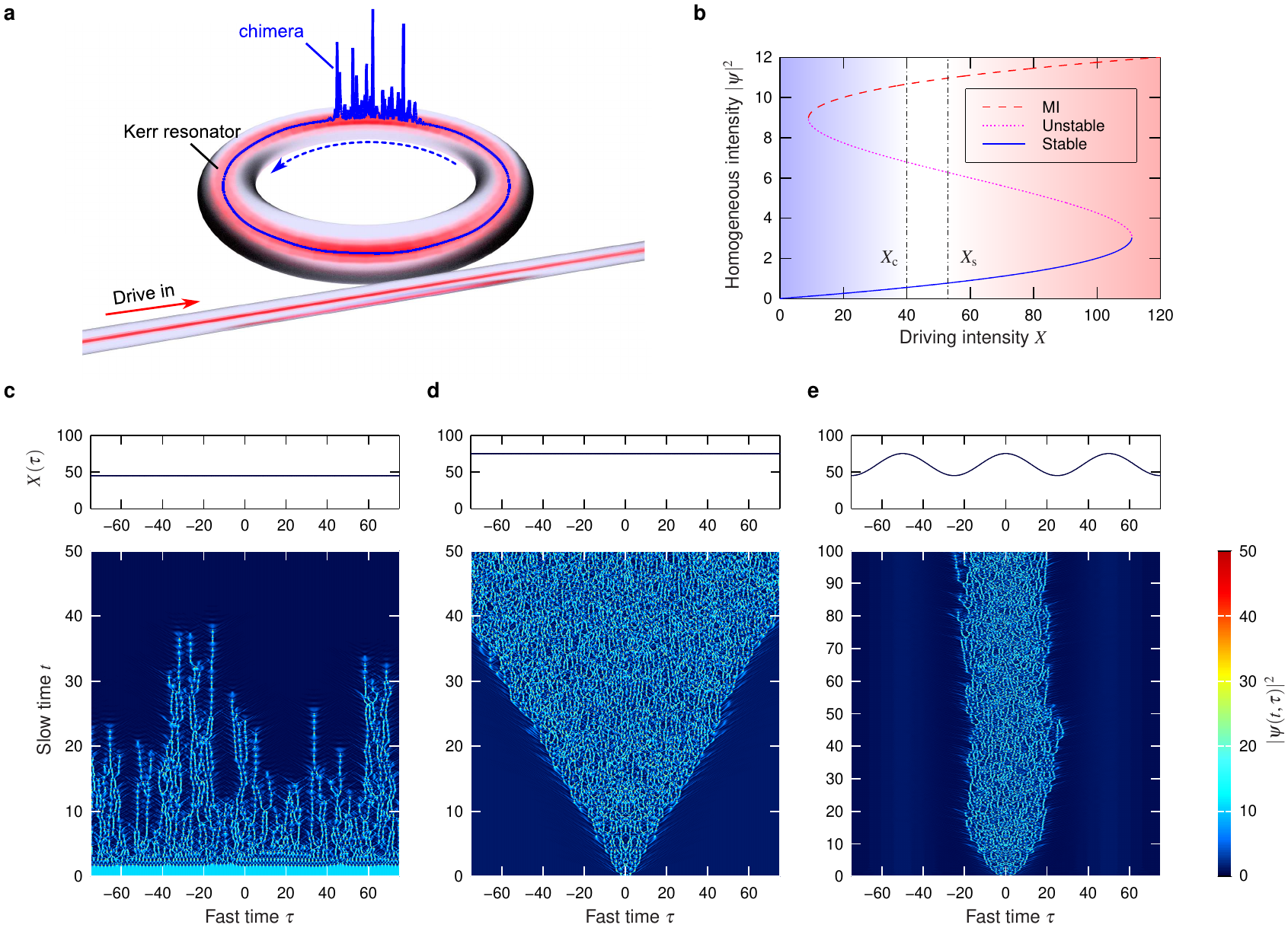}
  \caption{\textbf{Basic concept and illustrative simulation results.} \textbf{a}, In a dispersive Kerr resonator, chimera-like states correspond to localized chaotic domains that circulate around the resonator without expanding or contracting (on average). \textbf{b}, Intensity levels of the homogeneous steady-state solutions of Eq.~\eqref{LLN} for a detuning $\Delta = 9$ as a function of the driving intensity $X$. A spatiotemporally chaotic state can persist \emph{ad infinum} for driving intensities $X>X_\mathrm{s}\approx 53$ and decays to the homogeneous state for $X < X_\mathrm{s}$. For sufficiently small $X<X_\mathrm{c}\approx 40$, the decay is accompanied by the emergence of localized cavity solitons. \textbf{c, d}, Results from numerical simulations of Eq.~\eqref{LLN} with a homogeneous driving intensity of $X = 45$ and $X = 75$, respectively. \textbf{e}, Simulation results, showing how the expansion of a spatiotemporally chaotic domain can be arrested by a modulated driving field ($X_0 = 60$, $\omega = 2\pi/50$, $\varepsilon = 0.25$), thus giving rise to a persisting chimera-like state. Top panels in \textbf{c}--\textbf{e} show the driving intensity $X(\tau)$. All calculations and critical driving levels ($X_\mathrm{s}$ and $X_\mathrm{c}$) pertain to $\Delta = 9$. }
  \label{fig1}
\end{figure*}

\section{Illustrative numerical simulations}

Our system comprises of a dispersive, passive, Kerr nonlinear optical ring resonator that is coherently-driven with laser light. Here, the chimera-like states will correspond to temporally localized domains with complex and fluctuating internal structure that circulate around the resonator without contracting or expanding (on average) [see Fig.~\ref{fig1}(a)]. To describe the physics of such states, we consider the Lugiato-Lefever equation (LLE)~\cite{lugiato_spatial_1987} that provides the canonical description of the system under study~\cite{haelterman_dissipative_1992}. In a retarded frame of reference that is co-moving at the group velocity of light in the resonator, the LLE reads (see also Methods)
\begin{equation}
  \label{LLN}
  \frac{\partial \psi(t,\tau)}{\partial t} = \left[ -1 +i(|\psi|^2- \Delta)
  +i\frac{\partial^2}{\partial \tau^2}\right]\psi+\sqrt{X}.
\end{equation}
Here $\psi(t,\tau)$ describes the slowly-varying electric field envelope inside the resonator ($|\psi|^2$ is proportional to the experimentally accessible intensity), $t$ is a ``slow'' time variable that describes the evolution of $\psi(t,\tau)$ over consecutive roundtrips, and $\tau$ is a space-like variable that describes the envelope's profile over a single roundtrip. Physically, $\tau$ can be understood as the (angular) position in the resonator (in a co-rotating reference frame), or as a ``fast'' time that describes the relative temporal delay with which different portions of the intracavity envelope are coupled out of the resonator. The terms on the right-hand-side of Eq.~\eqref{LLN} respectively describe dissipation, Kerr nonlinearity, frequency detuning of the driving field from a cavity resonance ($\Delta$ is the detuning parameter), (anomalous) group-velocity dispersion, and coherent driving ($X$ is the driving intensity). Equation~\eqref{LLN} can be seen as a generic model of driven dissipative pattern formation close to the 1:1 resonance tongue~\cite{leo_dynamics_2013} with purely local (nearest neighbour), diffusion-like coupling (dispersion).

Chimera-like states require coexistence between a coherent and an incoherent state, which the system described by Eq.~\eqref{LLN} can provide at detunings $\Delta \gtrsim 4$ (ref.~\citen{leo_dynamics_2013}). In this regime, the homogeneous steady-state solutions of Eq.~\eqref{LLN} display an $S$-shape characteristic of bistability [Fig.~\ref{fig1}(b)], with the upper branch exhibiting a Turing-type modulation instability (MI) that gives rise to complex behaviour with spatiotemporally chaotic character~\cite{liu_characterization_2017,coulibaly_turbulence-induced_2019}. Numerical simulations of Eq.~\eqref{LLN} show that such behaviour can be sustained \emph{ad infinitum} only for driving intensities above a certain, detuning-dependent threshold ($X>X_\mathrm{s}$). For driving intensities below the threshold ($X<X_\mathrm{s}$), the spatiotemporally chaotic state exists only as a transient; it decays entirely to the homogeneous state [Fig.~\ref{fig1}(c)] for intermediate values ($X_\mathrm{c}<X<X_\mathrm{s}$) while persisting localized structures known as Kerr cavity solitons can emerge for lower values ($X<X_\mathrm{c}$). Such cavity solitons have drawn significant attention in the contexts of all-optical buffers~\cite{ackemann_chapter_2009, leo_temporal_2010} and ``Kerr'' microresonator optical frequency combs~\cite{kippenberg_dissipative_2018}, but do not play a role in our present study.

\begin{figure*}[!t]
\centering
  \includegraphics[width = \textwidth, clip = true]{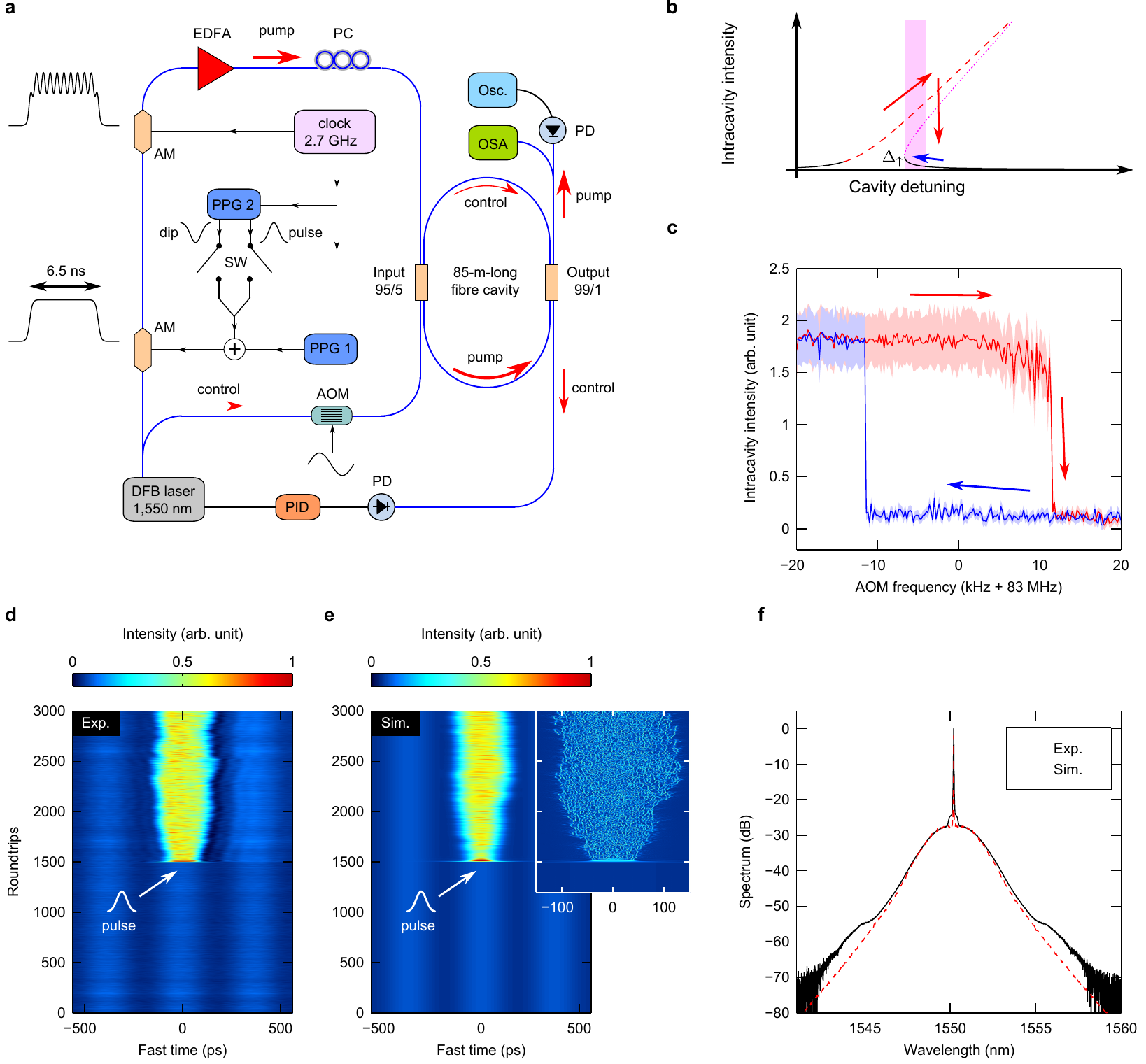}
  \caption{\textbf{Experimental concept and creation of chimera-like states.} \textbf{a}, Schematic diagram of the experimental setup. PPG: pulse pattern generator, SW: electronic switch, AM: amplitude modulator, EDFA: Erbium-doped fibre amplifier, PC: polarisation controller, AOM: acousto-optic modulator, PD: photodetector, Osc.: oscilloscope, OSA: optical spectrum analyser, PID: proportional-integral-derivative controller. \textbf{b}, Illustration of a nonlinearly tilted cavity resonance. Spatiotemporally chaotic and homogeneous states coexist for detunings slightly above the up-switching point $\Delta_\uparrow$ (magenta shaded region)~\cite{anderson_observations_2016}. \textbf{c}, Experimental measurement of the hysteresis around the up-switching point. Solid curves show the mean intensity of the intracavity field as the AOM frequency (hence, detuning) is increased (red curve) or decreased (blue curve). Shaded regions correspond to one quarter of the standard deviation of the recorded intensity, highlighting the incoherent (coherent) character of the upper (lower) state. \textbf{d}, Concatenation of oscilloscope traces, showing the excitation of a chimera-like state. \textbf{e}, Corresponding simulation results convolved with a 80~ps response function to mimic our experimental acquisition (inset shows results prior to convolution). \textbf{f}, Experimentally measured (black curve) and numerically simulated (red dashed curve) time-averaged spectrum of a single chimera-like state. }
  \label{fig2}
\end{figure*}

Whilst the coherent homogeneous state and the incoherent spatiotemporally chaotic state can both exist in quasi-steady-state when $X>X_\mathrm{s}$, the former is found to exhibit meta-stability at the expense of the latter. Specifically, a suitable perturbation that locally switches the stable homogeneous state to the modulationally unstable state results in the nucleation of an expanding domain of spatiotemporal chaos [Fig.~\ref{fig1}(d)]; the domain walls (fronts) that segregate the different states move with quasi-uniform velocity until the spatiotemporally chaotic state fills the entire system~\cite{anderson_observations_2016, pomeau_front_1986, liu_characterization_2017}. Because of the continuous and purely local nature of the system, it is not possible to find robust parameters for which the fronts would be motionless (on average), underlining the fact that discreteness underpins the emergence of chimera-like states in locally coupled systems~\cite{clerc_chimera-type_2016, clerc_chimera-like_2017}. On the other hand, discreteness from the perspective of continuous systems can be understood as a periodic potential akin to the Peierls-Nabarro potential of solid-state physics~\cite{clerc_continuous_2011}. Therefore, to arrest the front motion -- and hence realise persisting chimera-like states -- we \emph{engineer} discreteness in our system by applying a periodic modulation atop the cavity driving field, such that
\begin{equation}
X\rightarrow X(\tau) = X_0\left[1+\varepsilon \cos(\omega\tau)\right],
\label{modnorm}
\end{equation}
where $X_0$ is the average intensity of the drive, and $\varepsilon$ and $\omega$ are the depth and frequency of the modulation, respectively. Figure~\ref{fig1}(e) shows numerically simulated dynamics for such a drive [for parameters, see caption of Fig.~\ref{fig1}].  We see that a spatiotemporally complex domain initially expands, but motionless (on average) domain walls eventually form close to $\pm\tau_\mathrm{s}$, where the driving intensity $X(\tau_\mathrm{s}) = X_\mathrm{s}$. Further expansion of the domain is prohibited by the decay of perturbations entering the $X(\tau)<X_\mathrm{s}$ region [see Fig.~\ref{fig1}(c)], thereby allowing for a sustained chimera-like state of localized spatiotemporal chaos.

\section{Observation and control of chimera-like states}

For experimental demonstration, we use a macroscopic optical fibre ring resonator comprised of a 85-m-long segment of single-mode fibre that is looped on itself with a 99:5 coupler [see Fig.~\ref{fig2}(a) and Methods]. The resonator additionally includes a 99:1 coupler through which intracavity dynamics are monitored (i) in the time domain using a fast photodetector and an oscilloscope and (ii) in the spectral domain using an optical spectrum analyser.  We coherently drive the resonator with a train of flat-top, 6.5-ns-long pulses, obtained by amplitude modulating a distributed feedback fibre laser at 1550~nm with the help of a signal derived from a pulse pattern generator [PPG 1 in Fig.~\ref{fig2}(a)]. The PPG is referenced to an external clock at about 2.7~GHz, and we carefully adjust the clock frequency so as to synchronize the repetition rate of the driving pulse train with the round trip time of the resonator (about 420~ns). Using this same clock signal, we are able to synchronously imprint about 17 periods of sinusoidal modulation (with period of about 370~ps) atop the nanosecond pulses.  The modulation depth is set to $\varepsilon\approx 0.25$, and the nanosecond pulses are amplified to about 5.7~W (average power level across a single pulse, corresponding to $X_0 \approx 70$) before being launched into the cavity.

To experimentally identify the regime where spatiotemporally chaotic and homogeneous states coexist, we leverage the underlying hysteresis behaviour [Figs.~\ref{fig2}(b) and (c)]. Specifically, for fixed driving intensity, the coherent homogeneous and incoherent chaotic states coexist over a narrow region of cavity detunings just above the up-switching point $\Delta_\uparrow$ that marks the beginning of homogeneous bistability [Fig.~\ref{fig2}(b)]~\cite{leo_dynamics_2013,anderson_observations_2016}. To systematically access this regime, we use the detuning control scheme introduced in ref.~\citen{nielsen_invited_2018} (see also Methods): an acousto-optic modulator (AOM) is used to frequency shift a low-power, counter-rotating control beam whose intracavity intensity is locked at a set level. By adjusting the AOM frequency, we first stabilize the detuning at a value $\Delta < \Delta_\uparrow$, where the spatiotemporally chaotic state is the only state available. We then slowly increase the detuning lock-point (by adjusting the AOM frequency) so as to identify the value at which the chaotic state falls down to the homogeneous state. By slightly reducing the detuning from this value, we are guaranteed to be stabilized in the coexistence region.

Once the system is prepared, we deterministically \emph{excite} a chimera-like state. This is achieved by using a second pulse pattern generator [PPG 2 in Fig.~\ref{fig2}(a)] and an electronic switch to imprint a single perturbation pulse atop the modulated driving field (see Methods). Figure~\ref{fig2}(d) shows a concatenation of experimentally measured oscilloscope traces recorded at the 99:1 coupler of our cavity during an excitation event. The addressing pulse is launched into the cavity at round trip 1,500, and gives rise to a spatiotemporally complex domain that initially expands but remains confined within a single modulation cycle of the drive. Corresponding numerical simulations that use experimental parameters (see Methods) display similar behaviour [Fig.~\ref{fig2}(e)]. The simulations reveal that the chimera-like state consists of chaotically fluctuating pulsations with picosecond durations. Due to insufficient detector bandwidth, the experimental results shown in Fig.~\ref{fig2}(d) do not fully resolve the chimera's temporal profile; however, the time-averaged optical spectra of the experimentally measured and numerically simulated chimera-like states are in excellent agreement [Fig.~\ref{fig2}(f)], providing indirect evidence of similar internal structures.

\begin{figure}[!htb]
\centering
  \includegraphics[width = \columnwidth, clip = true]{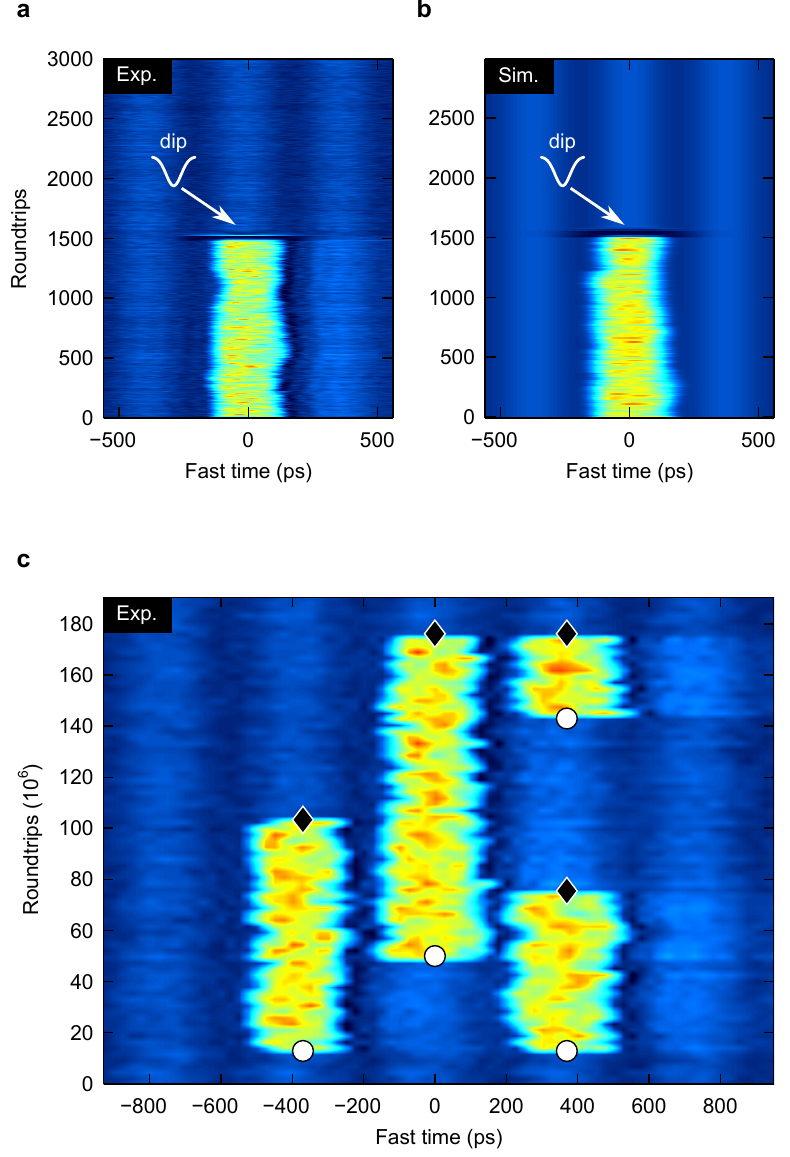}
  \caption{\textbf{Demonstration of erasure and manipulation of chimera-like states.} \textbf{a}, Concatenation of experimentally recorded oscilloscope traces, showing the erasure of a chimera-like state. \textbf{b}, Corresponding results from numerical simulations, convolved with an 80 ps detector response function to facilitate comparison with our experiments. \textbf{c}, Experimental demonstration of individual and parallel addressing of three chimera-like states. Solid white circles (black diamonds) highlight roundtrips and locations where pulse (dip) perturbations are added on the driving field. Colorbar for all panels is the same as in Figs.~\ref{fig2} (d) and (e).}
  \label{fig3}
\end{figure}

Once excited, the chimeras can persist for several minutes (limited only by our ability to maintain system stability), corresponding to tens of millions of characteristic cavity photon lifetimes. They can also be deterministically \emph{erased} by applying a localized dip on the driving field at the position of a chimera (see Methods). Figures~\ref{fig3}(a) and (b) show measured and simulated dynamics of such an erasure event, respectively. In contrast to excitation -- which can be achieved by means of just a single addressing pulse -- we have found that erasure generally requires the dip to be synchronously applied for several round trips. The excitation and erasure processes can also be applied in parallel, allowing for the individual and simultaneous addressing of several chimera-like states. Figure~\ref{fig3}(c) shows experimentally measured dynamics when three chimeras -- associated with three adjacent modulation cycles of the drive -- are turned on and off. As can be seen, the addressing of one state does not affect its neighbours. While similar individual addressing has been previously demonstrated for spatial~\cite{barland_cavity_2002} and temporal cavity solitons~\cite{leo_temporal_2010, wang_addressing_2018}, the results shown in Fig.~\ref{fig3} represent -- to the best of our knowledge -- the first demonstrations of such addressing for chimera-like states.

\begin{figure}[!htb]
\centering
  \includegraphics[width = \columnwidth, clip = true]{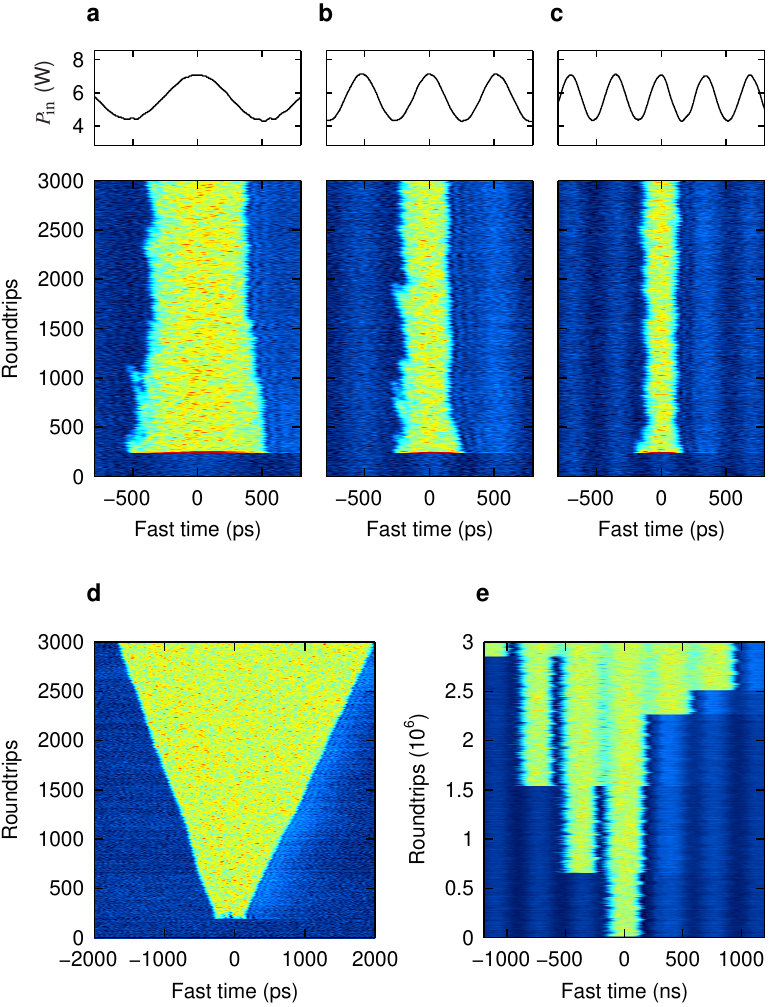}
  \caption{\textbf{Impact of modulation parameters.} \textbf{a}--\textbf{c}, Experimentally measured chimera excitation dynamics for three different modulation frequencies applied on the driving field: \textbf{a}, 0.96~GHz; \textbf{b}, 1.9~GHz; \textbf{c}, 2.9~GHz. For each case, the modulation depth $\varepsilon \approx 0.25$. The top panels show oscilloscope recordings of the driving field intensity profiles. \textbf{d} Quasi-linear expansion of a spatiotemporally chaotic domain observed in the absence of engineered discreteness ($\varepsilon = 0$).  \textbf{e}, Experimentally observed discrete transport of chimera-like states for an intermediate modulation depth $\varepsilon\approx 0.20$ and a modulation frequency of 2.7~GHz. Localized domains of spatiotemporal complexity undergo spontaneous depinning, intermittently expanding to their neighbouring modulation cycles. Colorbar for all panels is the same as in Figs.~\ref{fig2} (d) and (e). }
  \label{fig4}
\end{figure}

In addition to individual addressing, further control over the chimeras' characteristics and dynamics can be achieved by manipulating the granularity of the driving field. On the one hand, adjustment of the frequency of the modulation applied on the driving field offers control over the size of the spatiotemporally chaotic domains, as illustrated in Figs.~\ref{fig4}(a)--(c). Here we show experimentally measured chimera excitation dynamics for three different modulation frequencies as indicated; for each case, the incoherent domain remains localised over a single modulation cycle of the drive. Adjustment of the modulation depth on the other hand permits systematic exploration of the chimeras' localisation and transport dynamics. As the modulation depth is increased from zero, we observe a crossover from quasi-linear transport [Fig.~\ref{fig4}(d)] to complete localization of the chaotic state [c.f. Fig.~\ref{fig4}(c)]. Interestingly, for intermediate modulation depths, both our experiments and simulations show evidence of long-timescale discrete transport, where a chaotic domain remains localized for extended periods but eventually undergoes an abrupt expansion to fill the neighbouring cycle of the modulated drive [Fig.~\ref{fig4}(e)]. Such observations lead us to suppose that the fully localized chimeras in our system [c.f. Fig.~\ref{fig3}(c)] may correspond to transients with extremely long (and practically infinite) lifetimes, echoing the character of chimeras in oscillator networks with finite size~\cite{wolfrum_chimera_2011}.

\section{Time-lens imaging}

\begin{figure*}[!t]
\centering
  \includegraphics[width = \textwidth, clip = true]{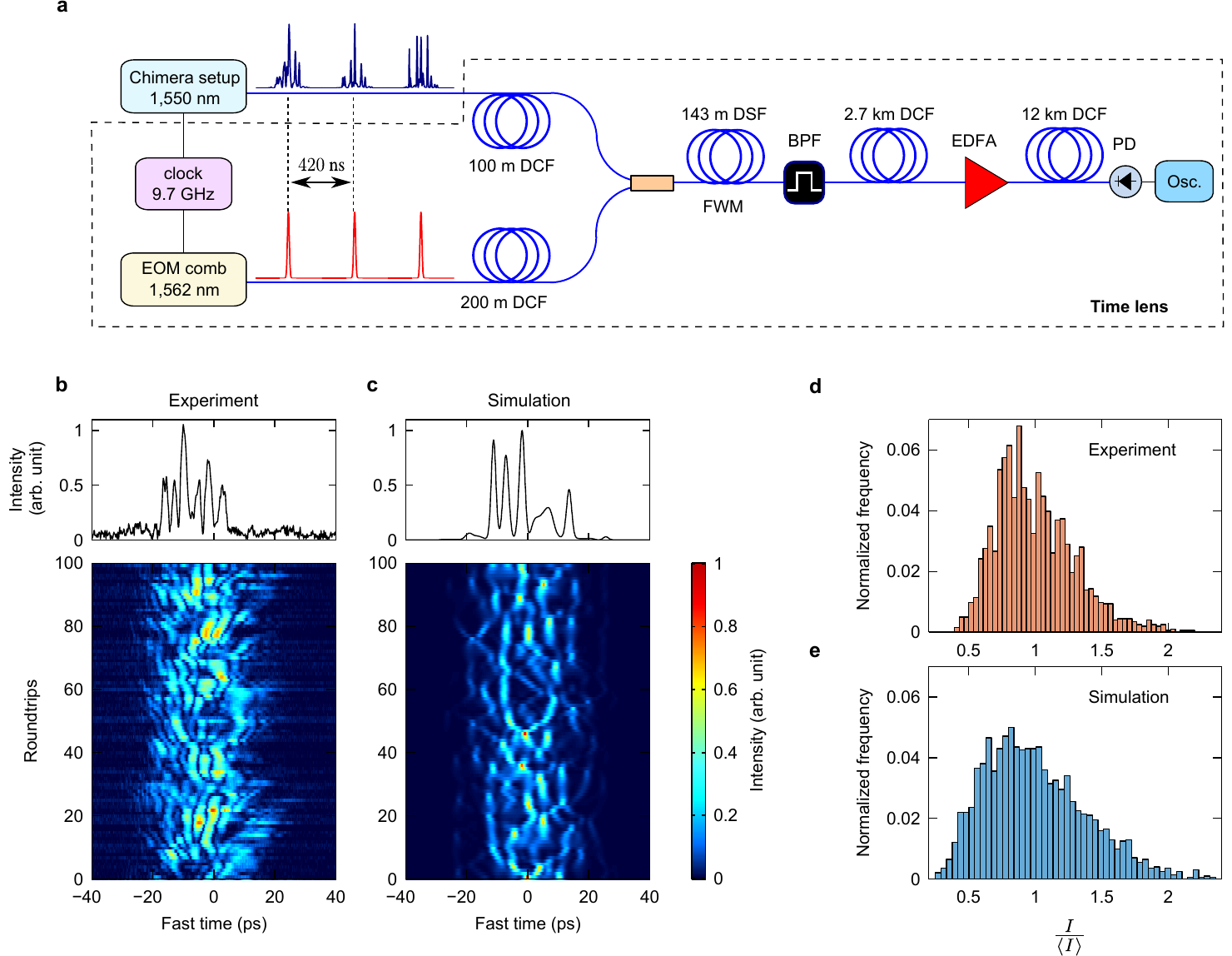}
  \caption{\textbf{Time lens imaging of picosecond-scale chimeras.} \textbf{a}, Schematic diagram of the experiment. EOM: electro-optic modulator, DCF: dispersion-compensating fibre, DSF: dispersion-shifted fibre, EDFA: Erbium-doped fibre amplifier, PD: photodetector, Osc.: oscilloscope. \textbf{b}, Experimental time lens results, showing the evolution of a chimera-like state over 100 round trips. \textbf{c}, Corresponding results from numerical simulations. The top panels in \textbf{b} and \textbf{c} show typical intensity profiles selected from the pseudo-color plots in more detail. \textbf{d, e}, Histograms showing the statistical distribution of the peak intensity $I$ of the intracavity field per round trip, normalized to the mean value $\langle I \rangle$. The statistics summarised in \textbf{d, e} encompass 2,000 roundtrips extracted from the same experimental (numerical simulation) results as \textbf{b} and \textbf{c}. To mimic the operation of our time lens, the simulation data has been filtered with a 25~ps temporal window centred around the chimera, and it has been convolved with a 1.5~ps Gaussian response function representative of the estimated time resolution of our system. }
  \label{fig5}
\end{figure*}
Because the chimeras in our system are anticipated to manifest themselves as non-repetitive, picosecond-scale pulsations [see inset of Fig.~\ref{fig2}(e)], they cannot be directly resolved using conventional time-domain experimental techniques. The experimental results shown in Figs.~\ref{fig2}--~\ref{fig4} are blurred by the 80~ps response time of our photodetectors, hiding the chimera's internal structure. To overcome this deficiency, we take advantage of recent advances in ultrafast optical metrology~\cite{suret_single-shot_2016,  narhi_real-time_2016, li_panoramic-reconstruction_2017, Ryczkowski_real-time_2018, copie_observation_2018}, and use a custom-built time-lens magnifier system with picosecond resolution to capture the round-trip-to-round-trip evolution of a chimera-like state.

Our time-lens system is based on the parametric four-wave mixing of a linearly-chirped pump pulse and the chimera state in a dispersion-shifted optical fibre [see Fig.~\ref{fig5}(a) and Methods]. It has a temporal resolution of about 1.5~ps, and is capable of capturing real-time dynamics over a temporal window of up to 25~ps. To reliably record the round-trip-to-round-trip evolution, the pump pulse must be precisely synchronized with the chimera. We achieve this by generating the pump with an electro-optic frequency comb setup that is driven at the same fundamental clock rate as the one used in the main chimera setup. With the help of additional delay lines, this synchronization ensures that the 25~ps temporal window always captures the same portion along the chimera. Up to 3,000 round-trips of evolution can be recorded at once, limited only by the memory depth of the oscilloscope used (see Methods).

Figure~\ref{fig5}(b) shows illustrative results from our time-lens experiments. We see clearly that the chimera is composed of picosecond-scale pulses that exhibit complex dynamics from roundtrip-to-roundtrip, in good agreement with corresponding numerical simulations [Fig.~\ref{fig5}(c)]. While the spatiotemporally chaotic nature of the dynamics hinders one-to-one comparison, we find that the experimentally measured statistical distribution of the roundtrip-to-roundtrip peak intensity is in good quantitative agreement with corresponding results from numerical simulations [Figs.~\ref{fig5}(d) and Figs.~\ref{fig5}(e)]. We must note that these results were obtained in the regime of intermediate modulation depth, where the spatiotemporally chaotic domains remain localised only for a finite number of round trips [see Fig.~\ref{fig4}(e)]. Unfortunately, due to the failure of a pulse pattern generator required for the experiment, we have not been able to repeat these measurements in the regime of full chimera localisation (see also Methods).

\section{Discussion}

In summary, we have experimentally shown that engineered discreteness allows for the persistence of chimera-like states in a continuous system with local coupling. Moreover, the system demonstrated in our work offers unprecedented control over the chimera's existence, characteristics, and dynamics. Compounded by the possibility to directly resolve the chimeras' internal structure and evolution in real time, our system offers a unique platform for experimental exploration of chimera physics. In this context, we must note that the ability to characterize non-repetitive ultrashort optical signals directly in the time domain has dramatically expanded our understanding of nonlinear instability dynamics over the past few years~\cite{suret_single-shot_2016,  narhi_real-time_2016, li_panoramic-reconstruction_2017, Ryczkowski_real-time_2018, copie_observation_2018}; our work extends such capabilities to the study of chimera-like states. Moreover, to the best of our knowledge, the results in Fig.~\ref{fig5}(b) represent the first direct roundtrip-by-roundtrip characterization of turbulent MI dynamics in coherently-driven Kerr resonators. Besides advancing our fundamental knowledge, better understanding of such dynamics may have applied implications for the design of frequency comb generators based on Kerr microresonators~\cite{karpov_dynamics_2019}.

\section*{Methods}

\small

\subparagraph*{\hskip-10pt Fibre resonator and driving field.} Our experiments were performed using two different fibre ring resonators made entirely of standard single-mode optical fibre (SMF-28). The results reported in Figs.~\ref{fig2}--\ref{fig4} were obtained using a resonator with a total length of about 85~m, corresponding to a measured free-spectral range of 2.39~MHz ($\pm 5~\mathrm{kHz}$) and a round trip time of 418~ns ($\pm 0.8~\mathrm{ns}$). The resonator includes 95/5 and 99/1 input and output couplers, respectively, and a polarization controller for adjusting the intracavity birefringence. The finesse of the resonator was measured to be 42 ($\pm 2$), corresponding to a 56.5~kHz ($\pm 2.7~\mathrm{kHz}$) resonance width. The time-lens results reported in Fig.~\ref{fig5} were obtained using a resonator similar to the one just described, but without an intracavity polarization controller. This resonator has a larger finesse of about 58, yielding a longer cavity photon lifetime that facilitates the real-time visualisation of the chimera-like states.

The resonator is driven by flat-top nanosecond pulses carved from a continuous wave (cw) distributed feedback fibre (DFB) laser at 1,550~nm (Koheras Adjustik E15), whose linewidth ($<1~\mathrm{kHz}$) is much smaller than the resonance width.  The reason for employing pulsed rather than cw driving is that the coexistence of spatiotemporally chaotic and homogeneous states requires a comparatively large intracavity intensity, which is difficult to achieve in cw~\cite{anderson_observations_2016}.  We must emphasize, however, that the chimeras studied in our work consist of picosecond-scale pulsations that are confined to domains with a duration of the order of 100~ps. Both of these numbers are significantly shorter than the flat-top portion of the nanosecond driving pulses, implying that the pulsed nature of the driving does not play a significant role in the chimera dynamics.

To create the nanosecond driving pulses, we modulate the intensity of the cw laser signal using a Mach-Zehnder amplitude modulator (AM) driven with a pulse pattern generator [PPG 1 in Fig.~\ref{fig2}(a)]. A sinusoidal waveform from an analog signal generator (Agilent Technologies) provides an RF clock signal to the PPG so as to precisely define the bit period of the PPG. The duration and separation of the nanosecond pulses then correspond to (different) integer multiples of this bit period. By carefully fine-tuning the frequency of the RF signal, we ensure that the driving pulse train is synchronised with the cavity. Note that, for experimental results shown in Fig.~\ref{fig2} and~\ref{fig3}, the RF clock signal was set to 2.7~GHz (bit period  370~ps), whilst the results shown in Fig.~\ref{fig4}(a) and (b) were obtained with lower frequencies as indicated. For the time-lens measurements shown in Fig.~\ref{fig5}, a larger frequency of 9.7~GHz (bit period 100~ps) was used. The need to use a larger frequency in the latter experiment was imposed by performance requirements of the time-lens system, which is referenced to the same RF clock signal as the chimera setup to ensure synchronization between the two (see also section below regarding the time-lens system).

After their generation, the nanosecond pulses pass through another Mach-Zehnder AM that imprints the sinusoidal modulation on top of them. This modulator is driven directly by the sinusoidal RF clock signal; accordingly, the frequency and period of the modulation -- and hence the size of the chimera domain -- is governed directly by the frequency of the RF clock. Before the driving pulses are launched into the resonator, they are amplified with an erbium-doped optical amplifier, and their polarization aligned with one of the two polarization eigenmodes of the resonator.

\subparagraph*{\hskip-10pt Detuning stabilization.} To actively stabilize and control the cavity detuning, we use the scheme introduced in Ref.~\citenum{nielsen_invited_2018}. A small portion of the driving laser beam is frequency-shifted by an acousto-optical modulator (AOM), and launched into the resonator such that it counter-rotates with respect to the main pump. One percent of the counter-rotating signal is extracted by the 99:1 coupler inside the resonator, and detected on a slow photodetector (response time much slower than the cavity round trip time). This detected signal is sent to a proportional-integral-derivative (PID) controller that fine-tunes the frequency of the driving laser via a piezo-electric actuator so as to maintain the detected signal at a set level. Changing the frequency of the AOM changes the detuning lock-point.

\subparagraph*{\hskip-10pt Chimera excitation and erasure.} To excite (erase) a chimera-like state, we add a pulse (dip) perturbation atop the driving field. This is achieved by using a second PPG [PPG 2 in Fig.~\ref{fig2}(a)] referenced to the same RF clock as the first one. The signal from the second PPG corresponds to a train of pulses or dips that repeat at the cavity roundtrip time, and an electronic switch controlled by a TTL signal is used to either let through or block the signal from the second PPG. When let through, the signal is electronically combined with the signal from the first PPG.  For chimera excitation, the TTL signal is adjusted to be shorter in duration than the cavity roundtrip time, such that only a single perturbation pulse is let through when the switch is open. For chimera erasure, the TTL signal is adjusted to be about $21~\mathrm{\mu s}$ long, such that a dip perturbation is synchronously added on the driving field for 50 round trips. Delay lines are used to ensure that the perturbations are applied at positions that correspond to the modulation maxima of the drive.

\subparagraph*{\hskip-10pt Detection.} We detect the intensity at the output port of the 99:1 coupler as a function of time using a fast 12.5~GHz photodetector, and digitise the detected signal with a~40~Gsamples/s real-time oscilloscope. Experimentally constructed spatiotemporal diagrams shown in Figs.~\ref{fig2}, Fig.~\ref{fig3}(a), Fig.~\ref{fig4}(a)--(d), and Fig.~\ref{fig5} were obtained by recording a single long time-trace with a total duration of about 1.25~ms (corresponding to about 3,000 roundtrips) at the oscilloscope's highest sampling rate.  The recorded signal is then split into segments that span a single round trip, and the resulting segments concatenated on top of each other to reveal the spatiotemporal dynamics. We must emphasize that these measurements capture the full roundtrip-by-roundtrip evolution of the intracavity field. However, the limited memory depth of the oscilloscope prevents recordings longer than 1.25~ms at the highest sampling rate. To demonstrate that the chimeras can persist (virtually) indefinitely, and to better capture their parallel control, the measurements shown in Fig.~\ref{fig3}(c) were obtained by recording a single oscilloscope trace per second over 75 seconds. This corresponds to 180 million round trips, or equivalently, about 28 million characteristic cavity photon lifetimes. The fact that the chimera-like states can persist over such timescale unless purposefully erased provides compelling evidence of their robustness. Finally, results shown in Fig.~\ref{fig4}(e) were obtained using the oscilloscope's segmented memory mode so as to capture one trace every 0.1~ms.

\subparagraph*{\hskip-10pt Time-lens measurements.} To generate the pump pulses in our time-lens system, cw light from an external cavity diode laser at 1563~nm is passed through an electro-optic modulation (EOM) comb generator that includes one amplitude modulator and two phase modulators. The modulators are driven by an RF clock signal at 9.7~GHz, which also provides the clock reference for the chimera setup. In the frequency domain, the EOM comb converts the cw signal into a comb of equally-spaced components spaced by 9.7~GHz. We must note that the spectral width of the resulting comb increases with the modulation frequency (with all else kept constant). Because this spectral width ultimately governs the bandwidth of the time-lens system, a large modulation frequency is preferred. It is for this reason that experiments reported in Fig.~\ref{fig5} were obtained with a larger RF frequency than those shown in Figa.~\ref{fig2}--~\ref{fig4}. Of course, the use of the 9.7~GHz clock signal in the main chimera setup requires two PPGs with bandwidth up to 10~GHz. Unfortunately, after performing the initial experiments shown in Fig.~\ref{fig5}, one of our 10~GHz PPGs broke, thus preventing time lens measurements in the fully localised chimera regime. Accordingly, comprehensive time-lens analysis of the chimeras' statistical properties is left for future work.

The output from the EOM comb generator must be dechirped to obtain a clean pulse train in the time-domain, which we achieve by passing the output through a 2.5-km-long segment of dispersion compensating fibre (DCF). An AOM driven by a third PPG referenced to the same RF clock as the rest of the setup is then used as a pulse picker to reduce the repetition rate of the resulting 9.7~GHz pulse train down to the repetition rate of the chimera cavity, i.e., about 2.4~MHz. The pulses are then amplified using an Erbium-doped fibre amplifier, followed by spectral broadening and corresponding temporal compression in a 1-km-long segment of SMF-28. This scheme allows us to generate a train of 4-ps-long pulses that are exactly synchronised with the nanosecond pump pulses used in the chimera setup. In contrast to other schemes, where an external mode-locked laser has been used as to generate the time-lens pump pulses~\cite{Ryczkowski_real-time_2018, copie_observation_2018}, our EOM comb approach ensures that the pump pulses are synchronised with the cavity dynamics under investigation for thousands of roundtrips.

Our time-lens system is designed to satisfy the thin-lens imaging condition~\cite{Ryczkowski_real-time_2018} $2/D_\mathrm{p} = 1/D_1 + 1/D_2$, where $D_\mathrm{p}$ and $D_1$ respectively represent the total group-velocity dispersions experienced by the pump and the chimera before they are nonlinearly mixed, and $D_2$ represents the total group velocity dispersion experienced by the nonlinear mixing product after its generation. More specifically, the pump pulses generated from the EOM comb setup described above are dispersed to 30~ps with about 200 metres of DCF ($D_\mathrm{p} = -22.3~\mathrm{ps/nm}$) and re-amplified with another EDFA. The output of the chimera setup is passed through about 100 metres of DCF ($D_\mathrm{1} = -11.13~\mathrm{ps/nm}$), combined with the pump pulses, and launched into a 143-m-long segment of dispersion-shifted fibre (DSF) with a zero-dispersion wavelength of 1561~nm. Four-wave-mixing between the pump and the chimera in the DSF gives rise a phase-conjugated replica of the chimera centred around 1576~nm, which we isolate using a spectral long-pass filter with a cut-off at 1565~nm.  The filtered signal is pre-stretched in one DCF module ($D_{21} = -333~\mathrm{ps/nm}$), amplified by an L-band EDFA, and passed through a second DCF module ($D_{22} = -1326~\mathrm{ps/nm}$), yielding a total dispersion of $D_\mathrm{2} = D_{21} + D_{22} = -1659~\mathrm{ps/nm}$. The signal is finally detected by a fast photodetector and a real-time oscilloscope. By dividing the time-base of the detected signal by the magnification factor of the time-lens system ($M=D_2/D_1\approx 150$), the space-time diagram in shown in Fig.~\ref{fig4}(c) is obtained.

\subparagraph*{\hskip-10pt Numerical model and simulations.} The dynamics of coherently-driven, passive, Kerr nonlinear resonators are well-known to be described by the Lugiato-Lefever equation (LLE)~\cite{leo_temporal_2010,anderson_observations_2016}. In dimensional form pertinent to the dispersive ring resonator considered in our experiments~\cite{haelterman_dissipative_1992}, the LLE reads
\begin{equation}
	t_R\frac{\partial E(t',\tau')}{\partial t'}=\left[ -\alpha-i(\delta_\mathrm{0}-\gamma L|E|^2)-\frac{iL\beta_2}{2}\frac{\partial^2}{\partial \tau'^2}\right]E+\sqrt{\theta}E_{\text{in}}.
\label{LLE}
\end{equation}
Here $t_\mathrm{R}$ is the round trip time of the resonator, $E(t',\tau')$ is the slowly-varying electric field envelope inside the resonator normalized such that $|E(t',\tau')|^2$ describes instantaneous power with units of Watts, $t'$ is the slow time variable that describes the field envelope's evolution over consecutive round trips, while $\tau'$ is a corresponding fast time variable defined in a reference frame that moves with the group velocity of light at the driving wavelength. $\alpha=\pi/\mathcal{F}$ with $\mathcal{F}$ the cavity finesse corresponds to half the total power loss per round trip with, $\delta_\mathrm{0}$ is the phase detuning of the driving field  from the closest cavity resonance, $L$ is the cavity round trip length, $\beta_2$ and $\gamma$ are the usual group velocity dispersion and Kerr nonlinearity coefficients, respectively, and $\theta$ is the power transmission coefficient of the coupler used to inject the driving field $E_\mathrm{in}$ with power $P_\mathrm{in} = |E_\mathrm{in}|^2$ into the resonator. We model the modulation of the driving field analogously with Eq.~\eqref{modnorm}:
\begin{equation}
P_\mathrm{in}(\tau') = P_0\left[1+\varepsilon \cos(\omega'\tau')\right],
\label{moddim}
\end{equation}
where $P_0$ is the average power of the drive and $\omega'$ is the frequency of modulation. The normalized Eq.~\eqref{LLN} is obtained by assuming anomalous dispersion ($\beta_2<0$) and defining the following dimensionless variables~\cite{leo_temporal_2010}: $t = \alpha t'/t_R$; $\tau = \sqrt{2\alpha/(|\beta_2|L)}\tau'$; $\psi = \sqrt{\gamma L/\alpha} E$; $\Delta = \delta_0/\alpha$;	$X = \gamma L \theta P_\mathrm{in}/\alpha^3$.

Simulations shown in Fig.~\ref{fig1} were obtained using the normalized Eq.~\eqref{LLN} with parameters quoted in the figure caption, whilst simulations shown in Figs.~\ref{fig2}--\ref{fig3} use Eq.~\eqref{LLE} with parameters estimated for the experiments: $\beta_2 = -20~\mathrm{ps^2/km}$; $\gamma = 1.2~\mathrm{W km^{-1}}$; $P_0 = 5.7~\mathrm{W}$; $L = 85~\mathrm{m}$; $\omega' = 2\pi\times 2.7~\mathrm{GHz}$, $\varepsilon = 0.25$; $\theta = 0.05$; $\mathcal{F} = 42$; $\delta_0 = 0.748~\mathrm{rad}$. Simulations shown in Fig.~\ref{fig5} likewise use the parameters estimated for that experiment:  $\beta_2 = -20~\mathrm{ps^2/km}$; $\gamma = 1.2~\mathrm{W km^{-1}}$; $P_0 = 2.7~\mathrm{W}$; $L = 85~\mathrm{m}$; $\omega' = 2\pi\times 9.7~\mathrm{GHz}$, $\varepsilon = 0.17$; $\theta = 0.05$; $\mathcal{F} = 58$; $\delta_0 = 0.596~\mathrm{rad}$.

All simulations use a split-step Fourier integration scheme with a time window that is equal to an integer multiple of the modulation period of the driving field. The initial condition for the simulations shown in Fig.~\ref{fig1}(c) corresponds to the upper homogeneous steady-state superposed with a broadband random noise seed to initiate modulation instability. The initial condition for the simulation shown in Fig.~\ref{fig1}(d) and (e) corresponds to a localized domain of the upper homogeneous state surrounded by the lower homogeneous state superposed with broadband random noise seed.

To simulate chimera excitation [Fig.~\ref{fig2}(e)], a Gaussian perturbation with 350~ps duration (full-width at half-maximum) and 50~W peak power is added on the driving power during a slow time interval that corresponds to a single roundtrip. A similar numerical scheme has previously been used to demonstrate excitation of cavity solitons~\cite{wang_addressing_2018}.

To simulate chimera erasure [Fig.~\ref{fig3}(c)], a Gaussian perturbation with 350~ps duration (full-width at half-maximum) and 7~W peak power is subtracted from the driving power, bringing the minimum driving power close to zero. As in our experiments, this perturbation is applied over 50 roundtrips.

\bibliographystyle{natphot}
\bibliography{Chimera}

\bigskip

\section*{Acknowledgments}

\noindent We acknowledge financial support from The Royal Society of New Zealand in the form of Rutherford Discovery (RDF-15-UOA-015, for M.E.) and James Cook (JCF-UOA1701, for S.C.) fellowships. We also thank Miles Anderson and Gang Xu for help in the initial and final stages of this project, respectively.

\section*{Author Contributions}

\noindent A.U.N. built the main chimera experiment and together with Y.X performed the chimera observation and manipulation experiments. Y.X. built the time-lens setup and performed the corresponding experiments. M.F., M.G.C, and S.C. helped with the theoretical interpretation of the results. M.F. and M.G.C. additionally provided background material relevant to the study. S.G.M. provided experimental support and advice. M.E. initiated the project with M.G.C., performed all numerical simulations, and supervised the project. M.E. wrote the manuscript with input from all the other authors.

\section*{Data availability}

\noindent The data that support the plots within this paper and other findings of this study are available from M.E. upon reasonable request.

\section*{Competing financial interests}

\noindent The authors declare no competing financial interests.

\end{document}